\documentclass[10pt,twocolumn]{article}

\usepackage[a4paper,top=18mm,bottom=20mm,left=16mm,right=16mm,columnsep=7mm]{geometry}
\usepackage{newtxtext,newtxmath}
\usepackage{microtype}
\usepackage{xcolor}
\usepackage{graphicx}
\usepackage{booktabs}
\usepackage{tabularx}
\usepackage{array}
\usepackage{tikz}
\usetikzlibrary{arrows.meta,positioning,fit,backgrounds}
\usepackage{pgfplots}
\pgfplotsset{compat=1.18}
\usepackage{hyperref}
\usepackage{xurl}
\usepackage{titlesec}
\usepackage{enumitem}
\usepackage{caption}
\usepackage{float}
\usepackage{placeins}
\usepackage{balance}
\usepackage{needspace}

\definecolor{vlteal}{HTML}{16615A}
\definecolor{vltealdark}{HTML}{0E4742}
\definecolor{vllight}{HTML}{E9F5F0}
\definecolor{vlpaper}{HTML}{FBFAF7}
\definecolor{vltext}{HTML}{16181D}
\definecolor{vlmuted}{HTML}{626A76}
\definecolor{vlrule}{HTML}{D8D4CB}
\definecolor{vlgold}{HTML}{A57A18}
\definecolor{vlred}{HTML}{B14949}

\hypersetup{
  colorlinks=true,
  linkcolor=vltealdark,
  citecolor=vltealdark,
  urlcolor=vlteal,
  pdftitle={Voice-Light: A Full-Duplex Cascaded Voice Agent with Causal Turn-Taking and Speculative Generation},
  pdfauthor={Bertil Braun}
}

\titleformat{\section}{\large\bfseries}{\thesection}{0.55em}{}
\titleformat{\subsection}{\normalsize\bfseries}{\thesubsection}{0.5em}{}
\titlespacing*{\section}{0pt}{1.15em}{0.45em}
\titlespacing*{\subsection}{0pt}{0.9em}{0.3em}

\setlist[itemize]{leftmargin=1.25em,itemsep=0.08em,topsep=0.25em}
\setlist[enumerate]{leftmargin=1.35em,itemsep=0.12em,topsep=0.25em}
\newcommand{\VoiceLight}{\textsc{Voice-Light}}
\newcommand{\code}[1]{\texttt{#1}}
\title{\textbf{Voice-Light: A Full-Duplex Cascaded Voice Agent with\\Causal Turn-Taking and Speculative Generation}}
\author{Bertil Braun\\
  \small \href{mailto:contact@bertil-braun.de}{contact@bertil-braun.de}}
\date{}

\begin{document}

\twocolumn[
\begin{@twocolumnfalse}
\maketitle
\vspace{-1.8em}
\begin{abstract}
Natural spoken interaction requires more than streaming ASR, language generation, and speech
synthesis: a system must react to overlap without canceling on every acknowledgment, prepare a
response before a turn is certain, and ensure canceled audio cannot enter conversation history. We
present \VoiceLight{}, a full-duplex cascaded voice agent that combines immediate acoustic onset,
a causal adapter sharing a streaming ASR encoder, reversible playback control, and private
speculative response generation. Structured tool calls execute concurrently with audible bridge
speech, while browser acknowledgments make rendered audio authoritative for durable history.
Locked evaluation on 1,673 real-conversation silence candidates found that an earlier learned
completion checkpoint preserved a 2.70\% false-cutoff rate but reached only 12.53\% end-of-turn
recall, compared with 95.60\% for a Silero timing policy. The deployed system therefore retains a
hybrid controller rather than claiming a learned-policy replacement. Across three unscripted
operator-run microphone sessions, 36 measured response turns had a 758 ms median from final VAD
endpoint to first server audio; 21 turns were below 800 ms. These sessions are an instrumented case
study, not a controlled user evaluation. We release the synthetic data, model artifacts, evaluation
code and summaries, source code, and deployment configuration supporting the result.
\end{abstract}
\vspace{0.45em}
\vspace{1.0em}
\end{@twocolumnfalse}
]

\section{Introduction}

A conventional voice assistant is often drawn as a serial pipeline: automatic speech recognition
(ASR), then a language model, then text-to-speech (TTS). That abstraction hides the interaction
problem. A usable conversational system must decide whether a silence is a completed turn or a
within-turn hesitation; react immediately when speech begins during playback without canceling on
every short acknowledgment; ensure canceled audio cannot reappear; and distinguish generated text
from speech that actually reached the listener.

Here, \emph{full duplex} means that microphone ingestion continues during assistant playback,
playback can react reversibly to detected user speech, and generation, synthesis, and acknowledged
audio are managed concurrently.

\VoiceLight{} retains a cascaded architecture because its boundaries expose typed tools, playback
state, and cancellation. Its central design rule is that uncertain work may begin early, but it may
become audible or durable only after explicit causal checks. This rule connects three otherwise
separate problems: turn inference must not inspect future audio, speculative generation must remain
private until promotion, and conversation history must contain only speech acknowledged by the
listener's browser.

The study asks three questions. First, can a small causal adapter reuse features from a persistent
streaming ASR encoder and improve turn commitment over deployable timing baselines? Second, can a
reversible controller exploit uncertain evidence without turning every VAD onset into cancellation?
Third, how much response latency can private speculation hide in a deployed ASR--LLM--TTS cascade?

The paper makes four scoped contributions:

\begin{itemize}
  \item a shared-encoder causal turn-taking adapter, together with locked evaluations that report
  both the learned signal and its failure to replace stronger timing baselines;
  \item a typed hybrid controller for reversible ducking, floor-taking interruption,
  backchannel resumption, stale-generation rejection, and audible-only history;
  \item private speculative generation integrated with streaming ASR, Qwen, Kyutai TTS, and
  sequential tool rounds in a public scale-to-zero deployment; and
  \item released synthetic tool-use and turn-taking corpora, trained artifacts, evaluation
  code and result summaries, and an instrumented 36-turn latency case study.
\end{itemize}

The contributions are systems and evaluation contributions. The paper does not claim that the
adapter is a new state of the art, that synthetic evaluation measures general tool competence, or
that the microphone sessions estimate population-level interaction quality.

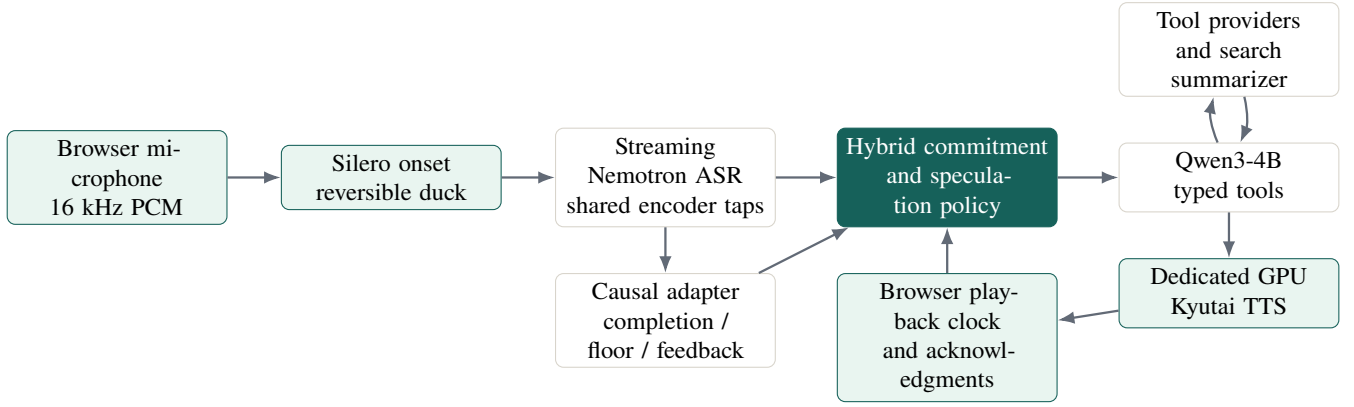
\begin{figure*}[t]
\centering
\resizebox{\textwidth}{!}{
\begin{tikzpicture}[
  node distance=0.65cm and 0.7cm,
  box/.style={draw=vlrule,rounded corners=3pt,fill=white,minimum height=0.75cm,text width=2.65cm,align=center,font=\small},
  green/.style={box,fill=vllight,draw=vlteal},
  dark/.style={box,fill=vlteal,draw=vlteal,text=white},
  arrow/.style={-{Latex[length=2.2mm]},line width=0.85pt,color=vlmuted}
]
  \node[green] (browser) {Browser microphone\\16 kHz PCM};
  \node[green,right=of browser] (silero) {Silero onset\\reversible duck};
  \node[box,right=of silero] (nemotron) {Streaming Nemotron ASR\\shared encoder taps};
  \node[box,below=0.6cm of nemotron] (adapter) {Causal adapter\\completion / floor / feedback};
  \node[dark,right=0.8cm of nemotron] (policy) {Hybrid commitment\\and speculation policy};
  \node[box,right=0.8cm of policy] (qwen) {Qwen3-4B\\typed tools};
  \node[green,below=0.6cm of qwen] (tts) {Dedicated GPU\\Kyutai TTS};
  \node[box,above=0.6cm of qwen] (tools) {Tool providers\\and search summarizer};
  \node[green,below=0.6cm of policy] (playback) {Browser playback clock\\and acknowledgments};
  \draw[arrow] (browser) -- (silero);
  \draw[arrow] (silero) -- (nemotron);
  \draw[arrow] (nemotron) -- (policy);
  \draw[arrow] (nemotron) -- (adapter);
  \draw[arrow] (adapter) -- (policy);
  \draw[arrow] (policy) -- (qwen);
  \draw[arrow] (qwen) to[bend left=18] (tools);
  \draw[arrow] (tools) to[bend left=18] (qwen);
  \draw[arrow] (qwen) -- (tts);
  \draw[arrow] (tts) -- (playback);
  \draw[arrow] (playback) -- (policy);
\end{tikzpicture}
}
\caption{Production runtime. Immediate acoustic reaction and learned interaction evidence meet at
a typed controller. Tool calls and results use a structured loop; acknowledged browser playback is
authoritative for durable assistant history.}
\label{fig:runtime}
\end{figure*}

\section{Related work}

\paragraph{Turn-taking and endpointing.}
TurnGPT predicts turn completion from text \cite{turngpt}; Voice Activity Projection (VAP) predicts
future speaker activity directly from conversational audio \cite{vap}; and acoustic--language-model
fusion shows that lexical and prosodic evidence can be complementary \cite{acousticllm}. Recent
turn benchmarks further emphasize that endpointing, backchannels, and interruption vary with
conversation type \cite{turnbench,fullduplexbench}. \VoiceLight{} does not propose a general
turn-taking architecture. It asks whether a small causal adapter can share the encoder already used
for streaming ASR and whether its evidence remains useful after a locked comparison with complete
deployable policies.

\paragraph{Full-duplex spoken agents.}
End-to-end systems such as Moshi jointly model user and assistant speech and avoid explicit
ASR--LLM--TTS boundaries \cite{moshi}. A cascade instead exposes intermediate text, typed tools,
playback state, and component failures. \VoiceLight{} uses those boundaries to make overlap
reversible, reject stale work, and construct history from acknowledged audio. Its latency results
are system measurements and are not directly comparable with model-internal latency reported for
end-to-end architectures. Speculative end-turn detection has also combined local and remote
detectors \cite{speculativeetd}; here, \emph{speculation} means privately preparing downstream text
and PCM before a final turn commitment.

\paragraph{Synthetic tool use.}
Toolformer and ToolLLM construct supervision for API selection and multi-step execution
\cite{toolformer,toollm}. The Voice-Light tool branch addresses a narrower spoken protocol: short
audible bridges, causally ordered call/result records, sequential calls, and suppression of
protocol markup. Its shared-generator holdout measures acquisition of that protocol, not general
tool use or factual retrieval.

\section{System overview}

Figure~\ref{fig:runtime} shows the deployed data path. The browser streams 16 kHz microphone PCM
over one WebSocket and acknowledges rendered audio ranges. Silero provides the earliest acoustic
onset signal. A persistent Nemotron worker produces streaming transcripts and exposes selected
encoder layers to the causal adapter without loading a second speech backbone. The controller
combines those signals with transcript revisions, playback state, and conservative deadlines.
Qwen generates speech and typed calls; Kyutai TTS \cite{kyutai} synthesizes only text released by
the controller.

Four invariants define the system. All learned turn features are causal and share the ASR stream.
Speech onset may duck or pause playback immediately, but only stronger lexical, learned, or timeout
evidence commits cancellation. Speculative text and PCM carry generation identifiers and remain
private until transcript and turn checks promote them. Finally, generated assistant text becomes
durable conversation context only to the extent that browser acknowledgments show it was audible.

\section{Training data}

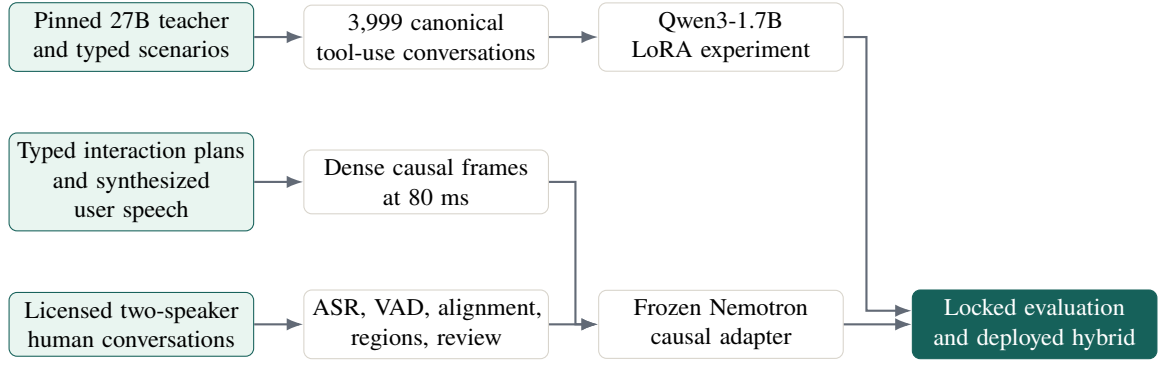
\begin{figure*}[t]
\centering
\begin{tikzpicture}[
  node distance=0.55cm and 0.65cm,
  box/.style={draw=vlrule,rounded corners=3pt,fill=white,minimum height=0.75cm,text width=3.0cm,align=center,font=\small},
  source/.style={box,fill=vllight,draw=vlteal},
  arrow/.style={-{Latex[length=2.2mm]},line width=0.8pt,color=vlmuted}
]
  \node[source] (toolsource) {Pinned 27B teacher\\and typed scenarios};
  \node[box,right=of toolsource] (tooldata) {3,999 canonical\\tool-use conversations};
  \node[box,right=of tooldata] (lora) {Qwen3-1.7B\\LoRA experiment};
  \node[source,below=0.8cm of toolsource] (audiosource) {Typed interaction plans\\and synthesized user speech};
  \node[box,right=of audiosource] (synthdata) {Dense causal frames\\at 80 ms};
  \node[source,below=0.8cm of audiosource] (human) {Licensed two-speaker\\human conversations};
  \node[box,right=of human] (annotation) {ASR, VAD, alignment,\\regions, review};
  \node[box,right=of annotation] (turn) {Frozen Nemotron\\causal adapter};
  \node[box,fill=vlteal,text=white,draw=vlteal,right=0.9cm of turn] (deploy) {Locked evaluation\\and deployed hybrid};
  \draw[arrow] (toolsource) -- (tooldata);
  \draw[arrow] (tooldata) -- (lora);
  \draw[arrow] (audiosource) -- (synthdata);
  \draw[arrow] (human) -- (annotation);
  \draw[arrow] (synthdata.east) -- ++(0.35,0) |- (turn.west);
  \draw[arrow] (annotation) -- (turn);
  \draw[arrow] (turn) -- (deploy);
  \draw[arrow] (lora.east) -- ++(0.3,0) |- ([xshift=-0.25cm,yshift=0.18cm]deploy.west) -- ([yshift=0.18cm]deploy.west);
\end{tikzpicture}
\caption{Two model-training branches feed one evaluated deployment. The turn-taking branch combines
synthetic interaction timelines with separately prepared human evidence; the language-model branch
uses synthetic typed conversations.}
\label{fig:pipeline}
\end{figure*}

The project produced two model-training branches. The language-model branch generated typed
spoken-tool conversations for a Qwen adapter. The turn-taking branch combined synthetic interaction
timelines with automatically prepared human conversations for a causal speech adapter. Figure
\ref{fig:pipeline} shows where the branches remain separate and where they meet the deployed system.

\subsection{Synthetic spoken tool use}

Spoken tool use was treated as a data and protocol problem rather than a prompt-only feature. A
pinned \code{Qwen3.6-27B-FP8} teacher generated provider-neutral records for natural
conversation, drafting, brainstorming, search, calculation, time queries, sequential calls, and
correction or recovery. The published corpus contains 3,999 conversations, 11,810 user messages, 15,308 assistant
messages, and 3,498 structured calls across \code{search}, \code{calculate}, and \code{get\_time}.
It is available through the project collection on Hugging Face \cite{tooldata}.

Generation used a causal conversation state machine. The teacher produced a user turn and then
either assistant speech or speech followed by a typed call. The controller executed or synthesized
the result, appended the linked call and result, and only then requested the next assistant step.
Calculator and time results were deterministic. Synthetic search results came from the teacher,
not the live web; consequently, the corpus teaches call ordering and grounding to supplied text,
not factual retrieval quality.

The canonical JSONL is semantic rather than tokenized. It stores runtime tool definitions, audible
assistant text, typed calls, stable identifiers, outcomes, later turns, split metadata, provenance,
and hashes. The run pins the teacher revision, prompt revision, source commit, scenario seeds,
quantization, and time anchor. Structural checks retained schema validity, causal call/result order,
deterministic tool checks, spoken-length limits, and protocol-leak detection. Subjective
teacher-based rejection was disabled after manual review found its false positives
counterproductive.

\subsection{Synthetic conversational turn-taking audio}

The second generator models coherent conversation timelines rather than isolated completion
utterances. Typed plans distinguish five semantic conditions: genuine user completion, a HOLD
pause followed by continuation, non-floor feedback during assistant speech, a response after the
assistant yields, and deliberate interruption. The plan also contains ordinary event-free user,
assistant, and silence spans.

Only user audio is synthesized. Virtual assistant text keeps the scenario coherent and estimates
assistant duration, but assistant waveform never becomes an input artifact. The runtime-relevant
assistant-speaking probability is compiled as a causal state curve with ramps, dips, and bounded
noise. This prevents future assistant state or clean synthetic timing from leaking into the
adapter.

The accepted rendering route used Qwen VoiceDesign references to condition CosyVoice 3 zero-shot
voices \cite{qwentts,cosyvoice}. Each user unit was rendered and trimmed independently;
the compiler then measured its active speech, assembled the final timeline, and rasterized
20-second views into 250 frames at 80 ms. A planned pause became HOLD supervision only if the
rendered unit contained at least 500 ms of silence followed by resumed speech.

The public synthetic repository \cite{turndata} stores lossless speech units and reconstruction metadata rather
than duplicating every composed waveform. Plans, voice provenance, trim measurements, composition
seeds, split assignments, and checksums allow conversation audio and training crops to be rebuilt.
The retained V4 and V5 manifests contain 1,092 and 1,097 timelines, approximately 21.84 and 21.80
hours of source timeline respectively. These counts describe retained run manifests, not a claim
that synthetic interaction validation substitutes for natural conversation.

\subsection{Automatic preparation of human conversations}

The human pipeline separates three artifact layers: authorized source audio, recording-level
evidence, and materialized training windows. Each complete speaker track is stored once as lossless
FLAC. Typed Parquet rows reference bounded 20-second intervals and contain aligned inputs, targets,
and masks; they do not duplicate the recording.

\Needspace{13\baselineskip}
Ingestion registers and hashes both channels, reuses exact cached transcripts, transports only
missing material, applies a reviewed inter-speaker offset, and persists a full-duration annotation
and quality result transactionally. Multiple ASR and activity systems provide independent evidence
for consensus alignment, channel-aware crosstalk filtering, conversation regions, and quality
flags. Supplied TurnBench annotations remain an independent evidence source rather than being
copied into model-output fields to manufacture agreement.

The completed preparation pass accepted eight MagicHub conversations (about 2.77 hours)
\cite{magichub} and 37 Mundo TurnBench conversations (about 7.31 hours) \cite{turnbench}; a silent
TurnBench recording was excluded.
These additions joined previously prepared authorized conversational material. The locked corpus
manifest records 107 accepted conversations totaling 36.3 hours across conversation-disjoint train,
validation, and test splits. A manual pipeline-validation audit reviewed 12 dataset-stratified recordings and
applied targeted exclusions; broader corpus quality was tracked through the automated evidence pipeline.

\begin{table}[t]
\centering
\caption{Human conversational material in the locked corpus manifest. Source audio remains subject
to its original access terms; only derived manifests and code are distributed by the project.}
\footnotesize
\begin{tabularx}{\columnwidth}{Xrr}
\toprule
Material & Conversations & Hours \\
\midrule
MagicHub & 8 & 2.77 \\
TurnBench & 37 & 7.31 \\
Additional authorized material & 62 & 26.22 \\
\midrule
Total & 107 & 36.30 \\
\bottomrule
\end{tabularx}
\label{tab:human-data}
\end{table}

\section{Model adaptation}

\subsection{A language adapter for spoken tools}

The tool-use experiment fine-tuned \code{Qwen/Qwen3-1.7B} \cite{qwen3} with a BF16 rank-16 LoRA \cite{lora}. Two source
records from each combination of eight behavior families and five speech styles formed a balanced
80-record validation split; the remaining 3,919 records were training data. Across 16 deterministic
passes, every record appeared once per pass. Intact short segments were regrouped into histories
with 8--16 user turns, producing 14,391 training histories and 17.84 million rendered tokens, of
which 6.86 million were assistant targets.

The native Qwen chat template rendered the canonical records with thinking disabled. The loss mask
retained ordinary assistant speech, the spoken bridge preceding a call, structured call tokens,
post-result continuation, and later assistant turns. System text, tool definitions, user messages,
tool results, padding, and non-assistant protocol tokens were masked. The adapter targeted all
attention and MLP projections, trained 17.43 million parameters, and completed 904 optimizer steps
in 47 minutes on an RTX 4090.

The public deployment uses Qwen3-4B Instruct \cite{qwen4}; the trained 1.7B LoRA remains a released
experimental artifact \cite{toollora,toolmerged}. Section~\ref{sec:offline-results} reports why the
synthetic protocol result was not treated as evidence of general conversational quality.

\subsection{A causal adapter on shared Nemotron features}

The turn-taking model attaches approximately 183,000 trainable parameters to the frozen
\code{Nemotron Speech Streaming 0.6B} encoder \cite{nemotron}. Nemotron is a cache-aware streaming
FastConformer-RNNT. Its activation caches allow new non-overlapping chunks to reuse prior work
\cite{fastconformer,stateful}. This property made it possible to share one streaming backbone
between ASR and turn inference rather than re-encoding a rolling waveform or loading a second
0.6-billion-parameter model.

The adapter taps encoder layers 6, 12, 18, and 24. Each 1,024-dimensional tap is normalized and
projected to 32 dimensions; the streams are fused to 64 dimensions, processed by residual causal
depthwise-separable convolutions, combined with assistant-speaking state, and passed through a
single-layer 64-dimensional unidirectional GRU. Separate heads emit completion/yield evidence,
interaction events including floor take and non-floor feedback, and future activity. Convolution
and recurrent state persist incrementally and reset with the ASR stream.

Synthetic pretraining selected step 2,250. Human fine-tuning warm-started those weights with a fresh
optimizer, 15\% synthetic replay, and 1,884 human boundaries: 1,038 HOLD and 846 end-of-turn (EOT)
examples. The resulting 750-step checkpoint was selected on human validation.

The final \code{adapter-best.pt} is step 750, with the pinned Nemotron revision listed in
Table~\ref{tab:repro}, one lookahead token, 80 ms encoder frames, and the four taps above. Human
validation selected it; synthetic validation could not. At runtime, forward
hooks capture the exact features produced by persistent streaming ASR. A bounded latest-value queue
supersedes stale adapter work, and loading or inference failure degrades the predictor without
stopping ASR or the voice session.

The offline protocols evaluate several checkpoints with different roles. Section~\ref{sec:offline-results}
keeps those roles explicit rather than transferring results between artifacts.

\section{Offline evaluation}
\label{sec:offline-results}

\subsection{Protocols and checkpoint lineage}

The project uses separate evaluation protocols for separate claims. Tool-use evaluation measures
behavior on held-out synthetic conversation states. Turn completion has two non-comparable
protocols: the original V1 benchmark scores future silence under a yield-oriented target and opens a
locked test split only after policy selection on validation, whereas V2 scores semantic completion
at speech boundaries. V2 stopped at validation because no detector met its predefined deployment
gate, so its test split remained sealed. Deployment evidence is an operator-run microphone case
study, not a substitute for a controlled conversational study.

\begin{table}[H]
\centering
\caption{Turn-adapter checkpoint lineage. Parenthesized values are training steps. ``Validation''
and ``test'' name distinct locked inventories; evidence is not transferred between artifacts.}
\footnotesize
\begin{tabularx}{\columnwidth}{@{}p{3.2cm}X@{}}
\toprule
Checkpoint (step) & Role and evidence \\
\midrule
Synthetic (2,250) & Pretraining seed; synthetic and human validation \\
Historical (3,500) & V1 locked test and V2 validation \\
Human fine-tune (750) & Deployed adapter; human validation and runtime only \\
Completion challenger (625) & Later experiment; V2 validation only \\
\bottomrule
\end{tabularx}
\label{tab:lineage}
\end{table}

\subsection{Synthetic tool-protocol results}

\begin{table}[t]
\centering
\caption{Balanced synthetic holdout for Qwen3-1.7B. The same 160 states were decoded with three
seeds. Expected calls, no-tool states, and continuations contribute 210, 90, and 180 repeated
predictions respectively; percentages therefore do not represent independent tasks.}
\scriptsize
\begin{tabularx}{\columnwidth}{Xrr}
\toprule
Behavior & Base & Fine-tuned \\
\midrule
Correct tool decision & 80.0\% & \textbf{95.2\%} \\
Exact tool name & 78.6\% & \textbf{94.3\%} \\
Valid argument schema & 80.0\% & \textbf{95.2\%} \\
Concise 1--12-word bridge & 7.6\% & \textbf{94.3\%} \\
Correctly avoided a tool & 70.0\% & \textbf{100.0\%} \\
Valid post-tool continuation & 98.3\% & \textbf{100.0\%} \\
\bottomrule
\end{tabularx}
\label{tab:tool}
\end{table}

The final adapter emitted 200/210 expected calls, 198/210 exact tool names, 200/210 schema-valid
arguments, 198/210 concise bridges, 90/90 correct no-tool responses, and 180/180 valid post-result
continuations. These descriptive gains demonstrate acquisition of the generator's conversational
protocol. They do not establish factual accuracy, robust web search, or general tool intelligence
because evaluation points share the training generator and schema.

\begin{figure}[t]
\centering
\begin{tikzpicture}
\begin{axis}[
  width=\columnwidth,height=4.2cm,
  xlabel={Completed training pass},ylabel={Validation loss},
  xmin=0.5,xmax=16.5,ymin=0.72,ymax=1.20,
  grid=major,grid style={vlrule!55},
  axis line style={vlmuted},tick style={vlmuted},
  label style={font=\small},tick label style={font=\small},
  line width=1.2pt
]
\addplot[color=vlteal,mark=*,mark size=1.8pt] coordinates {
 (1,0.9630) (2,0.8140) (3,0.7733) (4,0.7631) (5,0.7715)
 (6,0.7872) (7,0.8170) (8,0.8602) (9,0.9043) (10,0.9713) (11,1.0168)
 (12,1.0754) (13,1.1182) (14,1.1515) (15,1.1659) (16,1.1660)};
\addplot[color=vlgold,dashed] coordinates {(4,0.72) (4,1.20)};
\node[anchor=south west,font=\small,color=vlgold] at (axis cs:4.25,0.91) {loss minimum};
\end{axis}
\end{tikzpicture}
\caption{Recorded assistant-token validation loss after each logical pass. The directly comparable
zero-update loss was 4.5398 and is omitted to preserve resolution. Loss bottomed after pass 4, then
rose even while free-generation protocol metrics improved.}
\label{fig:loss}
\end{figure}
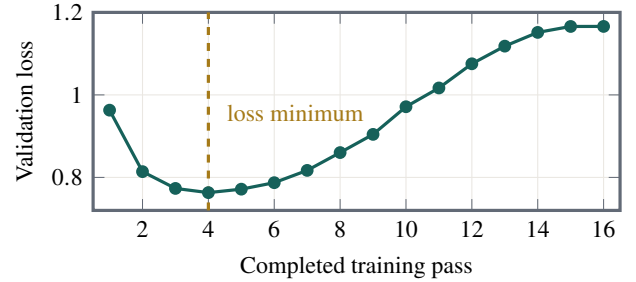

\subsection{Causal turn-completion results}

Synthetic pretraining exposed the transfer problem directly. Completion AUROC was 0.9260 on 893
held-out synthetic examples but 0.5646 on 789 clean human validation labels. Human fine-tuning
raised human AUROC to 0.5972. At its selected validation policy, the deployed step-750 checkpoint
reached 65.04\% EOT recall (426/655), 3.73\% false cutoffs (5/134 HOLD cases), and 2.0 s p95
commitment latency. These are validation results, not locked-test results, and the head remained a
poor standalone probability estimator.

For the policy metrics below, a HOLD candidate is a silence followed by continuation from the same
speaker, while EOT denotes a completed user turn. False cutoff is the fraction of HOLD candidates
committed as EOT; EOT recall counts completed turns committed before the timeout. Latency measures
the causal delay from the candidate boundary to commitment, capped by the earlier of the timeout or
the observed opportunity end.

The locked V1 test contains 1,673 causal silence candidates from 11 conversations, including only
37 HOLD cases. Threshold, minimum delay, and timeout were selected on validation and frozen before
test. Table~\ref{tab:v1} shows the central negative result: the Voice-Light checkpoint preserved a
low false-cutoff rate but crossed its learned threshold on only 205 of 1,636 EOT cases. Most turns
therefore used the 800 ms timeout, and the learned policy did not beat the Silero or LiveKit timing
baselines. This test evaluates historical step 3,500, not the deployed step-750 adapter.

The detectors do not score after identical amounts of silence. Voice-Light emits its boundary score
at 80 ms, Smart Turn at 240 ms, LiveKit at 320 ms, and Silero updates a silence proxy on 32 ms
chunks. Comparisons therefore evaluate complete causal policies at their native gates rather than
isolating model quality under identical acoustic evidence. The implementations are pinned to
Silero VAD 6.2.1, Pipecat Smart Turn v3.2, and LiveKit v1-mini \cite{silero,smartturn,livekit}.

\begin{table}[H]
\centering
\caption{Locked real-conversation test under the original V1 yield-target protocol. Mean latency
includes timeout actions. VL is historical Voice-Light step 3,500; policy values are threshold,
minimum delay, and timeout. Possible training-source overlap makes Smart Turn contextual only.}
\scriptsize
\setlength{\tabcolsep}{3.2pt}
\begin{tabularx}{\columnwidth}{Xrrrr}
\toprule
Metric & VL & Silero & Smart & LiveKit \\
\midrule
False cutoff & 2.70\% & 2.70\% & 13.51\% & 2.70\% \\
HOLD errors ($n=37$) & 1 & 1 & 5 & 1 \\
EOT recall & 12.53\% & 95.60\% & 20.72\% & 91.50\% \\
Mean latency & 770 ms & 656 ms & 684 ms & 654 ms \\
Threshold & 0.90 & 0.05 & 0.95 & 0.15 \\
Minimum delay & 560 ms & 640 ms & 80 ms & 640 ms \\
Timeout & 800 ms & 800 ms & 800 ms & 800 ms \\
\bottomrule
\end{tabularx}
\label{tab:v1}
\end{table}

The V2 semantic-completion validation inventory contains 1,005 soft targets and 789 clean hard
labels: 134 HOLD and 655 EOT examples, with 216 ambiguous cases excluded from hard scoring.
Historical step 3,500 achieved AUROC 0.6866, 65.80\% recall, 3.73\% false cutoffs, and 2.0 s p95
latency. Its BCE of 0.6719 and Brier score of 0.1843 were worse than the constant soft-prior
baseline (0.5847 and 0.1453), despite above-chance ranking. No swept detector met the joint gate of
at most 5\% false cutoff, at least 70\% recall, and at most 800 ms p95 latency, so the V2 test split
remained sealed. A completion-primary retrain improved calibration but still missed the gate;
longer training reduced recall.

\section{Streaming controller and deployment}

\subsection{Speculation and cancellation}

After 80 ms of low Silero speech probability, the adapter receives another 80 ms opportunity to
open a speculative response candidate. If learned evidence is not ready, a Silero fallback may
start at 160 ms. Candidate text and PCM remain private until final ASR and transcript-revision checks
permit promotion. Revisions limited to case, punctuation, whitespace, or apostrophes preserve work;
lexical change invalidates it.

The system uses generation identifiers, cancellation barriers, and rebased sample positions to
reject stale text or PCM. Playback clocks and boundary acknowledgements return every 80 ms. Durable
assistant history is constructed only from audio ranges that the browser confirms it rendered;
generated but canceled words remain diagnostic evidence rather than conversation content.

\subsection{Tool execution and speech boundaries}

Structured tools support calculation, current time, and web search, including sequential rounds.
Tool execution can overlap with bridge audio already buffered in the browser. Kyutai, however,
requires text lookahead. The bridge must therefore be closed before awaiting a tool result, and the
result begins a second TTS utterance within the same assistant generation. Reusing one uninterrupted
TTS session was implemented and reverted because it could strand the last bridge words until tool
completion, producing a deterministic mid-preamble stall. The final design avoids that stall but
can retain a short acoustic seam at the semantic boundary.
Before synthesis, a narrow text-boundary filter removes Markdown formatting, quotation marks, and
emoji that Kyutai would otherwise pronounce poorly; the durable generated text remains unchanged.

A separate Qwen3-0.6B worker \cite{qwen06} summarizes web results so provider output does not occupy the main
conversation worker. Tavily latency varied from approximately 234--408 ms in several calls to
roughly 1.98--2.29 seconds in slower observations. External provider time is therefore reported
separately from first-response latency.

\subsection{Hybrid overlap policy}

Silero remains the immediate acoustic authority. When user speech begins during assistant playback,
the browser fades toward -15 dB over 450 ms and pauses by 500 ms. This response is reversible: onset
alone does not discard generation. A floor-take probability above the configurable 0.82 threshold
can commit interruption and cancellation. A non-floor-feedback probability above 0.82 becomes
actionable after the short speech burst ends and resumes the same generation without adding a user
turn. Explicit stop, repair, question, or meaningful lexical evidence provides another fast path.

Ambiguous overlap is bounded. Sustained speech commits floor taking after 900 ms, while a short
ended burst resolves conservatively as non-floor feedback. Empty or slow final ASR receives a
120 ms grace rather than parking playback indefinitely. Generation and TTS budgets are held after
350 ms of unresolved overlap, and paused audio remains resumable for at most 800 ms. A committed
interruption rejects new server PCM immediately and fades no more than 100 ms of already-buffered
audio.

The adapter catches up from bounded pre-roll after Silero detects user activity, using the same
persistent Nemotron stream once active. This preserves one backbone and strict causality while
limiting the acoustic context available at the earliest overlap decision.

\subsection{Two GPUs and scale to zero}

Production requests a co-located pair of A10 GPUs, with an L40S pair as the capacity fallback.
GPU 0 runs Nemotron, the adapter, Qwen3-4B, and
the small search summarizer. GPU 1 is reserved for Kyutai. Moving Nemotron beside Kyutai improved
Qwen first-delta latency by roughly 25--30 ms but regressed Kyutai first-word-to-PCM in small warm
traces: 346--353 ms with dedicated TTS ($n=3$) versus 459--486 ms in the alternate placement
($n=5$). Because first audible speech dominates perceived response time, the dedicated-TTS
topology was retained; the traces are engineering placement checks rather than statistical tests.

Modal \cite{modal} is a thin wrapper around the provider-neutral compute application. Model revisions and the
adapter are pinned; Hugging Face and Torch caches live on a persistent volume; serving operates
offline after initialization. Model subprocesses persist inside a warm container.

The final topology loads Nemotron, Qwen, Kyutai, and the search summarizer concurrently after the
small VAD is ready. Cached weights prevent downloads but not image scheduling, imports, CUDA
initialization, or framework startup. Observed cold readiness remains approximately 32--55 seconds.
GPU snapshots failed for the multi-process CUDA layout, and an import-only CPU snapshot added about
30 seconds. Keeping a container warm was rejected for the expected low visit frequency and cost.

\section{End-to-end latency case study}

The deployment case study comprises three unscripted microphone sessions run by one operator from
a browser in Germany against the scale-to-zero service in Modal's European region. Measurements
begin after the service reported readiness and therefore exclude cold start. Prompts covered
ordinary conversation, longer responses, interruptions, short acknowledgments, calculations,
current information, and sequential tool use. The final VAD endpoint is the last speech boundary
used for turn commitment; first server audio is the first PCM packet sent for the response.

Across 36 measured response turns, the median final-VAD-to-first-server-audio latency was 758 ms;
21/36 turns were below 800 ms, and the range was 528--1,652 ms. One session exported the complete
telemetry trace for 13 turns. In that subset, median endpoint commitment was 502 ms, LLM first word
336 ms after generation start, TTS first PCM 444 ms after its first word, and server-to-browser
render 95 ms. These component intervals overlap and must not be added.

Nine of the 13 fully traced turns promoted speculative work. Their median final-VAD-to-first-server-
audio latency was 667 ms, compared with 1,513 ms for the four turns without promotion. This
observational split is consistent with speculation hiding downstream work, but candidate success is
confounded with transcript stability and turn difficulty; it is not a causal estimate of an 846 ms
speedup. Kyutai's fixed first-frame computation remained the largest warm downstream cost.

\begin{figure*}[t]
\centering
\includegraphics[width=0.96\textwidth]{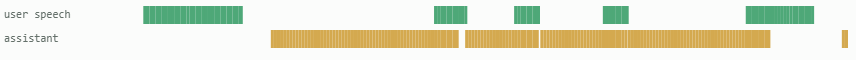}
\caption{Observable activity during a 20-second excerpt from an interactive microphone session.
Green spans show detected user speech and gold spans show acknowledged assistant playback. The
session exercised short backchannels and a floor-taking interruption, but the lanes intentionally do
not assign semantic labels to individual overlaps.}
\label{fig:debug}
\end{figure*}

\begin{table}[H]
\centering
\caption{Deployment case-study measurements. Aggregate response latency covers three sessions;
component medians come from the 13-turn fully exported trace and overlap in time.}
\begin{tabularx}{\columnwidth}{Xr}
\toprule
Measurement & Observed value \\
\midrule
Sessions / measured turns & 3 / 36 \\
Final VAD to first server audio, median & \textbf{758 ms} \\
Final VAD to first server audio, mean & 932 ms \\
Final VAD to first server audio, range & 528--1,652 ms \\
Turns below 800 ms & 21 / 36 \\
Fully traced turns & 13 \\
Final VAD to endpoint commitment, median & 502 ms \\
Generation start to LLM first word, median & 336 ms \\
TTS first word to first PCM, median & 444 ms \\
Server audio send to browser render, median & 95 ms \\
Final VAD to browser render, median & 836 ms \\
\bottomrule
\end{tabularx}
\label{tab:latency}
\end{table}

The original 250--600 ms engineering target was not met consistently. The 800 ms line became a
practical development reference, not a population target. The sessions were neither scripted nor
blinded, involved one operator, and omit users, acoustic conditions, accents, and networks outside
the development setting. They do not justify population p50 or p95 claims or independently quantify
interruption, duck, or backchannel-resume latency.

\section{Observability and reproducibility}

The browser's diagnostic stream separates model evidence from durable conversation. Its rolling
20-second view records user-speech and assistant-audible lanes, raw adapter outputs, Silero state,
applicability, causal audio time, age, inference latency, policy decisions, and browser
acknowledgments. The adapter heads were not independently calibrated as probabilities of observable
conversational behavior; their values are therefore retained as engineering telemetry rather than
presented as behavioral results. Figure~\ref{fig:debug} limits the paper view to directly observable
speech and playback activity.

Concrete reproducibility artifacts accompany each boundary: immutable serialized schemas reject
extra fields; manifests retain seeds, input hashes, configuration, metrics, and checkpoint identity;
locked evaluation files carry content hashes and detector provenance; and build and test commands
are recorded with pinned model revisions. Deployment packages the final adapter and mounts
persistent model caches without committing credentials. Table~\ref{tab:repro} identifies the final
runtime snapshot and principal pins.

The public artifacts include the tool-use dataset, LoRA, merged 1.7B checkpoint, and synthetic
turn-taking audio \cite{tooldata,toollora,toolmerged,turndata}. Human conversational audio remains
restricted by its source terms. The source repository contains generation, preparation, training,
and evaluation code; narrative result summaries; the runtime and browser; and Modal deployment
runbooks. Large prediction inventories and restricted source audio are not public artifacts.

\FloatBarrier
\section{Discussion and limitations}

The principal result is not a learned endpointer that replaces silence timing. It is a bounded way
to integrate uncertain learned evidence into a streaming controller. Immediate onset remains
acoustic, early reactions are reversible, private work is discarded safely, and a deadline provides
progress when the learned model is late or unavailable. This separation allowed the adapter to
remain an optional signal after its standalone completion policy failed the locked gate.

The experiments also expose two transfer failures. Synthetic turn pretraining produced strong
in-domain ranking but transferred weakly to human completion labels. Synthetic tool fine-tuning
improved a shared-generator protocol holdout but did not establish general reasoning or retrieval
quality. Qwen3-4B was selected for deployment after an engineering integration review, not a
controlled model-quality comparison. Together, these results argue for measuring learned
components at the boundary where they will be used rather than inferring deployment value from
training-domain metrics.

Several limitations constrain the evidence. The locked V1 turn test contains 11 conversations and
only 37 HOLD cases, so one error changes false cutoff by 2.70 percentage points and candidates within
a conversation are correlated. Possible training-source overlap prevents a clean Smart Turn
superiority comparison. V2 excludes 216 ambiguous cases and lacks the planned independent human
label audit. Most importantly, the deployed step-750 adapter was not evaluated on the locked V1
test, and its floor-take and backchannel heads have not been calibrated against an independently
annotated natural-conversation test set.

The deployment study adds only three unscripted sessions from one operator. It is not a user study
and does not establish naturalness, interruption accuracy, or latency distributions across users,
accents, noise, echo, networks, or GPU types. The system is English-first, human audio remains
restricted by source terms, and factual generation depends on a small language model and external
retrieval. A stronger evaluation would use conversation-disjoint human labels, participant-level
confidence intervals, controlled overlap scenarios, and final-topology action latencies. A learned
policy should replace the hybrid controller only if it improves the latency-versus-false-cancel
frontier on that evidence.

\section{Conclusion}

\VoiceLight{} demonstrates an artifact-backed path from causal conversational data to a public
full-duplex cascaded agent \cite{demo,voice-light}. The system shares streaming ASR features with a
small turn adapter, prepares responses speculatively, executes typed tools, controls playback
reversibly, and records only acknowledged audio as durable assistant history. Its three-session
case study shows sub-second median server response while retaining scale-to-zero deployment.

The negative result is equally central: the historical learned completion policy did not beat
simple deployable timing baselines on the locked real-conversation test. Instrumentation therefore
changed the design rather than being used to justify the original model. The resulting contribution
is a reproducible hybrid system and a set of explicit evidence boundaries for future controlled
turn-taking evaluation.

\balance
\begingroup
\small
\interlinepenalty=10000

\endgroup

\clearpage
\onecolumn
\appendix
\section{Reproducibility snapshot}

\begin{table}[H]
\centering
\caption{Principal revisions, hashes, and recorded validation for the final runtime. Full commands
and additional hashes are recorded in the repository runbooks \cite{voice-light}.}
\footnotesize
\begin{tabularx}{\textwidth}{@{}p{4.1cm}X@{}}
\toprule
Artifact & Revision, hash, or recorded result \\
\midrule
Report source & Committed TeX, figure assets, and build instructions in the source repository \\
Runtime source snapshot & \code{e2f79adc} on \code{master}; evaluated two-GPU deployment \code{d78115d2} \\
Qwen3-4B-Instruct-2507 & \nolinkurl{cdbee75f17c01a7cc42f958dc650907174af0554} \\
Qwen3-0.6B summarizer & \nolinkurl{c1899de289a04d12100db370d81485cdf75e47ca} \\
Nemotron Streaming 0.6B & \nolinkurl{ebe59e5a817142986528bbbee5dba8db7b38ed50} \\
Kyutai TTS 1.6B & \nolinkurl{f65439609986c392cb12df63938abcc550c3fb15} \\
Deployed adapter step 750 & SHA-256 \nolinkurl{d5c8e02c61dc9c230eac57992278383b81f14e3b71935b8dc684fe5da4171011} \\
Release engineering checks & 630 Python tests (1 skipped, 1 integration deselected), 38 browser tests, Ruff clean; not scientific evaluation \\
\bottomrule
\end{tabularx}
\label{tab:repro}
\end{table}

\subsection{Build and validation commands}

The release PDF is built from the committed source. From the repository root on Windows, the
recorded validation sequence is:

\begin{small}
\begin{verbatim}
.\.venv\Scripts\ruff.exe format --check --exclude .cache .
.\.venv\Scripts\ruff.exe check --exclude .cache .
.\.venv\Scripts\python.exe -m pytest tests\voice_agent tests\training\turn_taking
    -m "not integration" --import-mode=importlib
node --test tests\browser\*.test.mjs
tectonic --outdir output\pdf docs\technical-report\voice-light-technical-report.tex
\end{verbatim}
\end{small}

Model-dependent integration tests and the deployed microphone acceptance session require the
documented external services, GPU environment, and uncommitted credentials; they are not implied by
the local command sequence above.

\end{document}